\documentstyle[amssymb,aps,12pt]{revtex}

\begin{document}
\draft
\author{Sergio De Filippo\cite{byline}}
\address{Dipartimento di Fisica ''E. R. Caianiello'', Universit\`{a} di Salerno\\
Via Allende I-84081 Baronissi (SA) ITALY\\
Tel: +39 089 965 229, Fax: +39 089 965 275, e-mail: defilippo@sa.infn.it\\
and \\
Unit\`{a} INFM Salerno}
\date{\today}
\title{The Schroedinger-Newton model as $N\rightarrow \infty $ limit of a $N$ color
model.}
\maketitle

\begin{abstract}
The generalization to $N$\ colors of a recently proposed non unitary two
color model for the gravitational interaction in non relativistic quantum
mechanics is considered. The $N\rightarrow \infty $ limit is proven to be
equivalent to the Schroedinger-Newton model, which, though sharing
localization properties with the $N=2$ model, cannot produce decoherence.
\end{abstract}


In some recent papers\cite{defilippo1,defilippo2,defilippo3} a model for
gravitational interactions in nonrelativistic quantum mechanics was
proposed. Starting from a gravity-free generic system, all matter degrees of
freedom were duplicated and gravitational interactions were introduced
between observable (green) and unobservable (red) degrees of freedom only.
Once unobservable degrees of freedom are traced out, the ensuing non unitary
dynamics includes both the traditional aspects of classical gravitational
interactions and a form of fundamental decoherence, which may be relevant to
the emergence of the classical behavior of macroscopic bodies. The model in
fact treats on an equal footing mutual and self-interactions, which, by some
authors, are possibly held responsible for wave function localization and/or
reduction \cite{karolyhazy,diosi0,diosi,penrose}.

In particular, for bodies of ordinary density, there is a localization
threshold at about $10^{11}$ proton masses and linear superpositions of two
localized wave packets may decohere, either due to fast phase oscillations 
\cite{defilippo1}, or due to the slow spreading of the center of mass
probability density. Above the localization threshold, the latter gives
rise, even starting from a single localized wave packet, to the emergence of
delocalized ensembles of localized wave packets, whose entropy is slowly
growing in time\cite{defilippo2}. While the model was originally presented
in terms of instantaneous action at a distance gravitational interactions,
in Ref.\cite{defilippo3} it was shown that a Hubbard-Stratonovich
transformation, together with the replacement of the instantaneous
interaction by a retarded one, leads to a field-theoretic interpretation.
The presence of both a positive and a negative energy scalar field implies
the complete cancellation among all Feynman diagrams containing divergent
self-energy or vertex insertions, making the theory finite, without any mass
or coupling constant renormalization\cite{defilippo3}.

As to the relationship between our model and Einstein gravity, the viewpoint
we adhered to is that the latter may be presumably only a large scale
manifestation of a fundamental theory that may well be out of reach, and
whose possible non relativistic limit is the object of our proposal\cite
{defilippo1}. More specifically it was hinted that the Einstein theory could
arise, unlike, for instance, classical electrodynamics, not as a result of
taking expectation values with respect to a pure physical state, but rather,
as an effective long distance theory like hydrodynamics, from a statistical
average, or equivalently by tracing out unobservable degrees of freedom from
a pure metastate \cite{defilippo3}. A similar point of view about gravity as
''...an emergent phenomenon, in the same sense that fluid dynamics emerges
from molecular physics...'' is also suggested by analog models of general
relativity, \ \ based on condensed matter physics and leading to consider
scalar fields as possible starting points\cite{barcelo}.

It was also suggested\cite{defilippo3} that a suitable mean field
approximation gives the Schroedinger-Newton (SN) model\cite
{christian,penrose1}, which is a nonlinear generalization of the
Schroedinger equation, \ where gravitational self-interactions give rise to
stationary localized wave functions\cite{moroz,tod,kumar,melko}, contrary to
the spreading mixed states of our model\cite{defilippo2}. More generally the
SN approximation cannot produce decoherence, so that, in order to estimate
localization times, due to the lack of a full theory of quantum gravity, one
has to invoke purely dimensional arguments\cite{penrose,melko}. On the
contrary, according to the exact model, time evolution of an unlocalized
state, represented as a linear superposition of many localized wave packets,
can lead explicitly to vanishing coherences and then to an ensemble of such
localized states \cite{defilippo3}.

The aim of the present paper is to give a well defined procedure for passing
from the original model to the SN approximation, replacing the heuristic
suggestion of Ref.\cite{defilippo3}. In doing that, the original model with
just two colors, green and red, is considered as the simplest
representative, for $N=2$, of a class of $N$-color models, whereas the SN
model is obtained as the $N\rightarrow \infty $ limit. While clarifying the
relationship between our proposal and the SN model, this result gives in
principle the possibility to develop $1/N$ expansions in analogy to what is
done in ordinary condensed matter physics\cite{stanley,largeN}. In
particular, while the $N\rightarrow \infty $ limit does not involve
decoherence, but only localization, $1/N$ expansions may provide approximate
schemes for the evaluation of decoherence.

To be specific, following Ref. \cite{defilippo1}, let $H[\psi ^{\dagger
},\psi ]$ denote the second quantized non-relativistic Hamiltonian of a
finite number of particle species, like electrons, nuclei, ions, atoms
and/or molecules, according to the energy scale. For notational simplicity $%
\psi ^{\dagger },\psi $ denote the whole set $\psi _{j}^{\dagger }(x),\psi
_{j}(x)$ of creation-annihilation operators, i.e. one couple per particle
species and spin component. This Hamiltonian includes the usual
electromagnetic interactions accounted for in atomic and molecular physics.
To incorporate gravitational interactions including self-interactions, we
introduce a color quantum number $\alpha =1,2,...,N$, in such a way that
each couple $\psi _{j}^{\dagger }(x),\psi _{j}(x)$ is replaced by $N$
couples $\psi _{j,\alpha }^{\dagger }(x),\psi _{j,\alpha }(x)$ of
creation-annihilation operators. The overall Hamiltonian, including
gravitational interactions and acting on the tensor product $%
\bigotimes_{\alpha }F_{\alpha }$ of the Fock spaces of the $\psi _{\alpha }$
operators, is then given by 
\begin{equation}
H_{G}=\sum_{\alpha =1}^{N}H[\psi _{\alpha }^{\dagger },\psi _{\alpha }]- 
\frac{G}{N-1}\sum_{j,k}m_{j}m_{k}\sum_{\alpha <\beta }\int dxdy\frac{\psi
_{j,\alpha }^{\dagger }(x)\psi _{j,\alpha }(x)\psi _{k,\beta }^{\dagger
}(y)\psi _{k,\beta }(y)}{|x-y|},
\end{equation}
where here and henceforth Greek indices denote color indices, $\psi _{\alpha
}\equiv (\psi _{1,\alpha }\psi _{2,\alpha },...\psi _{N,\alpha })$, and $%
m_{i}$ denotes the mass of the $i$-th particle species, while $G$ is the
gravitational constant.

While the $\psi _{\alpha }$ operators obey the same statistics as the
original operators $\psi $, we take advantage of the arbitrariness
pertaining to distinct operators and, for simplicity, we choose them
commuting with one another: $\alpha \neq \beta \Rightarrow $ $[\psi _{\alpha
},\psi _{\beta }]$ $_{-}=[\psi _{\alpha },\psi _{\beta }^{\dagger }]_{-}=0$.
The metaparticle state space $S$ is identified with the subspace of $%
\bigotimes_{\alpha }F_{\alpha }$ including the metastates obtained from the
vacuum $\left| \left| 0\right\rangle \right\rangle =\bigotimes_{\alpha
}\left| 0\right\rangle _{\alpha }$ by applying operators built in terms of
the products $\prod_{\alpha =1}^{N}\psi _{j,\alpha }^{\dagger }(x_{\alpha })$
and symmetrical with respect to arbitrary permutations of the color indices,
which, as a consequence, for each particle species, have the same number of
metaparticles of each color. This is a consistent definition since the time
evolution generated by the overall Hamiltonian is a group of (unitary)
endomorphisms of $S$. If we prepare a pure $n$-particle state, represented
in the original setting - excluding gravitational interactions - by 
\begin{equation}
\left| g\right\rangle \doteq \int d^{n}xg(x_{1},x_{2},...,x_{n})\psi
_{j_{1}}^{\dagger }(x_{1})\psi _{j_{2}}^{\dagger }(x_{2})...\psi
_{j_{n}}^{\dagger }(x_{n})\left| 0\right\rangle ,
\end{equation}
its representative in $S$ is given by the metastate 
\begin{equation}
\left| \left| g^{\otimes N}\right\rangle \right\rangle \doteq \prod_{\alpha
} \left[ \int d^{n}xg(x_{1},x_{2},...,x_{n})\psi _{j_{1},\alpha }^{\dagger
}(x_{1})...\psi _{j_{n},\alpha }^{\dagger }(x_{n})\right] \left| \left|
0\right\rangle \right\rangle .  \label{initial}
\end{equation}
As for the physical algebra, it is identified with the operator algebra of
say the $\alpha =1$ metaworld. In view of this, expectation values can be
evaluated by preliminarily tracing out the unobservable operators, namely
with $\alpha >1$, and then taking the average of an operator belonging to
the physical algebra. It should be made clear that we are not prescribing an
ad hoc restriction of the observable algebra. Once the constraint
restricting $\bigotimes_{\alpha }F_{\alpha }$ to $S$ is taken into account,
in order to get an effective gravitational interaction between particles of
one and the same color\cite{defilippo1}, the resulting state space does not
contain states that can distinguish between operators of different color.
The only way to accommodate a faithful representation of the physical
algebra within the metastate space is to restrict the algebra.

While we are talking trivialities as to an initial metastate like in Eq. (%
\ref{initial}), that is not the case in the course of time, since the
overall Hamiltonian produces entanglement between metaworlds of different
color, leading, once unobservable operators are traced out, to mixed states
of the physical algebra. It was shown in Ref. \cite{defilippo1} that, for $%
N=2$, the ensuing non-unitary evolution induces both an effective
interaction reproducing gravitation, and wave function localization. The
proof of the former property stays conceptually unchanged for arbitrary $N$
and is then omitted here. A peculiar feature of the model is that it cannot
be obtained by quantizing its naive classical version, since the classical
states corresponding to the constraint in $\bigotimes_{\alpha }F_{\alpha }$,
selecting the metastate space $S$, have partners of all colors sitting in
one and the same space point and then a divergent gravitational energy.
While it is usual that, in passing from the classical to the quantum
description, self-energy divergences are mitigated, in this instance we pass
from a completely meaningless classical theory to a quite divergence free
one. This is more transparent in a field theoretic description\cite
{defilippo3}.

Let us adopt here an interaction representation, where the free Hamiltonian
is identified with $\sum_{\alpha =1}^{N}H[\psi _{\alpha }^{\dagger },\psi
_{\alpha }]$ and the time evolution of an initially unentangled metastate $%
\left| \left| \tilde{\Phi}(0)\right\rangle \right\rangle =\bigotimes_{\alpha
=1}^{N}\left| \Phi (0)\right\rangle _{\alpha }$\ is represented by 
\begin{eqnarray}
\left| \left| \tilde{\Phi}(t)\right\rangle \right\rangle &=&{\it T}\exp %
\left[ \frac{iG}{(N-1)\hslash }m^{2}\sum_{\alpha <\beta }\int dt\int dxdy%
\frac{\psi _{\alpha }^{\dagger }(x,t)\psi _{\alpha }(x,t)\psi _{\beta
}^{\dagger }(y,t)\psi _{\beta }(y,t)}{|x-y|}\right] \left| \left| \tilde{\Phi%
}(0)\right\rangle \right\rangle  \nonumber \\
&\equiv &U(t)\left| \left| \tilde{\Phi}(0)\right\rangle \right\rangle ,
\label{evolvedmetastate}
\end{eqnarray}
where for notational simplicity we are referring to just one particle
species.

Then, by using a Stratonovich-Hubbard transformation\cite{negele}, we can
rewrite $U(t)$ as 
\begin{eqnarray}
U(t) &=&\int {\it D}\left[ \varphi \right] \prod_{\alpha }{\it D}\left[
\varphi _{\alpha }\right] \exp \frac{ic^{2}}{2\hslash }\int dtdx\left[
\varphi (x,t)\nabla ^{2}\varphi (x,t)-\sum_{\alpha }\varphi _{\alpha
}(x,t)\nabla ^{2}\varphi _{\alpha }(x,t)\right]  \nonumber \\
&&{\it T}\exp \left[ -i\frac{2mc}{\hslash }\sqrt{\frac{\pi G}{N-1}}
\sum_{\alpha }\int dtdx\left[ \varphi (x,t)+\varphi _{\alpha }(x,t)\right]
\psi _{\alpha }^{\dagger }(x,t)\psi _{\alpha }(x,t)\right] ,
\label{stratonovich}
\end{eqnarray}
i.e. as a functional integral over the auxiliary real scalar fields $\varphi
,\varphi _{1},\varphi _{2},...,\varphi _{N}$, which, though less economical,
for $N=2$, than that in Ref.\cite{defilippo3}, is the simplest one for a
generic $N$.

Just as in Ref.\cite{defilippo3}, a physical interpretation of this result
can be given, by considering the minimal variant of the Newton interaction
in Eq. (\ref{evolvedmetastate}) aiming at avoiding instantaneous action at a
distance, namely replacing $-1/|x-y|$ by the Feynman propagator $4\pi
\square ^{-1}\equiv 4\pi \left( -\partial _{t}^{2}/c^{2}+\nabla ^{2}\right)
^{-1}$. Then the analog of Eq. (\ref{stratonovich}) holds with the
d'Alembertian $\square $ replacing the Laplacian $\nabla ^{2}$ and the
ensuing expression can be read as the mixed path integral and operator
expression for the evolution operator corresponding to the field Hamiltonian 
\begin{eqnarray}
H_{Field} &=&\sum_{\alpha =1}^{N}H[\psi _{\alpha }^{\dagger },\psi _{\alpha
}]+\frac{1}{2}\int dx\left[ \pi ^{2}+c^{2}\left| \nabla \varphi \right|
^{2}-\sum_{\alpha =1}^{N}\left( \pi _{\alpha }^{2}+c^{2}\left| \nabla
\varphi _{\alpha }\right| ^{2}\right) \right]  \nonumber \\
&&+2mc\sqrt{\frac{\pi G}{N-1}}\int dx\sum_{\alpha =1}^{N}\left\{ \left[
\varphi +\varphi _{\alpha }\right] \psi _{\alpha }^{\dagger }\psi _{\alpha
}\right\} ,
\end{eqnarray}
where $\pi =\dot{\varphi}$ and $\pi _{\alpha }=\dot{\varphi}_{\alpha }$
respectively denote the conjugate fields of $\varphi $ and $\varphi _{\alpha
}$ and all fields are quantum operators. This can be read in analogy with
nonrelativistic quantum electrodynamics, where a relativistic field is
coupled with nonrelativistic matter, while the procedure to obtain the
corresponding action at a distance theory by integrating out the $\varphi $
fields is the analog of the Feynman's elimination of electromagnetic field
variables\cite{feynman}.

The resulting theory, containing the negative energy fields $\varphi
_{\alpha }$, has the attractive feature of being divergence free, at least
in the non-relativistic limit, where Feynman graphs with virtual
particle-antiparticle pairs can be omitted. To be specific, it does not
require the infinite self-energy subtraction needed for instance in
electrodynamics on evaluating the Lamb shift, or the coupling constant
renormalization\cite{itzykson}. Here of course we refer to the covariant
perturbative formalism applied to our model, where matter fields are
replaced by their relativistic counterparts and the non-relativistic
character of the model is reflected in the mass density being considered as
a scalar coupled with the scalar fields by Yukawa-like interactions. In fact
there is a complete cancellation, for a fixed color index $\alpha $, among
all Feynman diagrams containing only $\psi _{\alpha }$ and internal $\varphi 
$ and $\varphi _{\alpha }$ lines, owing to the difference in sign between
the $\varphi $ and the $\varphi _{\alpha }$ free propagators. This state of
affairs of course is the field theoretic counterpart of the absence of
direct $\psi _{\alpha }-\psi _{\alpha }$ interactions in the theory obtained
by integrating out the $\varphi $ operators, whose presence would otherwise
require the infinite self-energy subtraction corresponding to normal
ordering.

Going back to our metastate (\ref{evolvedmetastate}), the partial trace 
\begin{equation}
M(t)\equiv \stackrel{\circ }{Tr}\left| \left| \tilde{\Phi}(t)\right\rangle
\right\rangle \left\langle \left\langle \tilde{\Phi}(t)\right| \right|
\equiv \sum_{k_{2},k_{3},...,k_{N}}\;\;\;\left\langle \left\langle
k_{2},k_{3},...,k_{N}\right| \right| \left| \left| \tilde{\Phi}
(t)\right\rangle \right\rangle \left\langle \left\langle \tilde{\Phi}
(t)\right| \right| \left| \left| k_{2},k_{3},...,k_{N}\right\rangle
\right\rangle ,
\end{equation}
gives the corresponding physical state, where $\left| \left|
k_{2},k_{3},...,k_{N}\right\rangle \right\rangle \equiv \bigotimes_{\alpha
=2}^{N}\left| k_{\alpha }\right\rangle _{\alpha }$ and $(\left|
k\right\rangle _{\alpha })_{k=1,2,...}$ denotes an orthonormal basis in the
Fock space $F_{\alpha }$. Then, by using Eq. (\ref{stratonovich}), we can
write 
\[
M(t)= 
\]
\begin{eqnarray}
&&\int {\it D}\left[ \varphi \right] \prod_{\alpha }{\it D}\left[ \varphi
_{\alpha }\right] {\it D}\left[ \varphi ^{\prime }\right] \prod_{\alpha } 
{\it D}\left[ \varphi _{\alpha }^{\prime }\right] \exp \frac{ic^{2}}{
2\hslash }\int dtdx\left[ \varphi \nabla ^{2}\varphi -\sum_{\alpha }\varphi
_{\alpha }\nabla ^{2}\varphi _{\alpha }-\varphi ^{\prime }\nabla ^{2}\varphi
^{\prime }+\sum_{\alpha }\varphi _{\alpha }^{\prime }\nabla ^{2}\varphi
_{\alpha }^{\prime }\right]  \nonumber \\
&&\left[ \bigotimes_{\alpha =2}^{N}\;_{\alpha }\left\langle \Phi (0)\right| %
\right] {\it T}^{-1}\exp \left[ i\frac{2mc}{\hslash }\sqrt{\frac{\pi G}{N-1}}
\sum_{\alpha =2}^{N}\int dtdx\left[ \varphi ^{\prime }(x,t)+\varphi _{\alpha
}^{\prime }(x,t)\right] \psi _{\alpha }^{\dagger }(x,t)\psi _{\alpha }(x,t) %
\right]  \nonumber \\
&&{\it T}\exp \left[ -i\frac{2mc}{\hslash }\sqrt{\frac{\pi G}{N-1}}
\sum_{\alpha =2}^{N}\int dtdx\left[ \varphi (x,t)+\varphi _{\alpha }(x,t) %
\right] \psi _{\alpha }^{\dagger }(x,t)\psi _{\alpha }(x,t)\right]
\bigotimes_{\alpha =2}^{N}\left| \Phi (0)\right\rangle _{\alpha }  \nonumber
\\
&&{\it T}\exp \left[ -i\frac{2mc}{\hslash }\sqrt{\frac{\pi G}{N-1}}\int dtdx %
\left[ \varphi (x,t)+\varphi _{1}(x,t)\right] \psi _{1}^{\dagger }(x,t)\psi
_{1}(x,t)\right] \left| \Phi (0)\right\rangle _{1}  \nonumber \\
&&_{1}\left\langle \Phi (0)\right| {\it T}^{-1}\exp \left[ i\frac{2mc}{
\hslash }\sqrt{\frac{\pi G}{N-1}}\int dtdx\left[ \varphi ^{\prime
}(x,t)+\varphi _{1}^{\prime }(x,t)\right] \psi _{1}^{\dagger }(x,t)\psi
_{1}(x,t)\right] .
\end{eqnarray}

To study the $N\rightarrow \infty $ limit of this expression, replace the
products $\varphi \psi _{1}^{\dagger }\psi _{1}$ and $\varphi ^{\prime }\psi
_{1}^{\dagger }\psi _{1}$ respectively with $\tilde{\varphi}\psi
_{1}^{\dagger }\psi _{1}$ and $\tilde{\varphi}^{\prime }\psi _{1}^{\dagger
}\psi _{1}$, inserting simultaneously the $\delta $-functionals $\delta
\lbrack \varphi -\tilde{\varphi}]=\int {\it D}\left[ g\right] \exp (i\frac{%
2mc}{\hslash }\sqrt{\frac{\pi G}{N-1}}\int dtdxg[\tilde{\varphi}-\varphi ])$%
, $\delta \lbrack \varphi ^{\prime }-\tilde{\varphi}^{\prime }]=\int {\it D}%
\left[ g^{\prime }\right] \exp (i\frac{2mc}{\hslash }\sqrt{\frac{\pi G}{N-1}}%
\int dtdxg^{\prime }[\varphi ^{\prime }-\tilde{\varphi}^{\prime }])$. Then
we can perform functional integration on $\varphi $,$\varphi ^{\prime }$ and 
$\varphi _{\alpha }$,$\varphi _{\alpha }^{\prime }$ for $\alpha =2,3,...,N$
and get 
\[
M(t)= 
\]
\begin{eqnarray}
&&\int {\it D}\left[ g\right] {\it D}\left[ g^{\prime }\right] \left[
\bigotimes_{\alpha =2}^{N}\;_{\alpha }\left\langle \Phi (0)\right| \right] 
\nonumber \\
&&{\it T}^{-1}\exp \frac{-iG}{\hslash }m^{2}\int dt\int dxdy\left[
\sum_{1<\alpha <\beta }\frac{\psi _{\alpha }^{\dagger }(x)\psi _{\alpha
}(x)\psi _{\beta }^{\dagger }(y)\psi _{\beta }(y)}{(N-1)|x-y|}+\frac{\left[
\sum_{1<\alpha }\psi _{\alpha }^{\dagger }(x)\psi _{\alpha }(x)+g^{\prime
}(x)/2\right] g^{\prime }(y)}{(N-1)|x-y|}\right]  \nonumber \\
&&{\it T}\exp \frac{iG}{\hslash }m^{2}\int dt\int dxdy\left[ \sum_{1<\alpha
<\beta }\frac{\psi _{\alpha }^{\dagger }(x)\psi _{\alpha }(x)\psi _{\beta
}^{\dagger }(y)\psi _{\beta }(y)}{(N-1)|x-y|}+\frac{\left[ \sum_{1<\alpha
}\psi _{\alpha }^{\dagger }(x)\psi _{\alpha }(x)+g(x)/2\right] g(y)}{%
(N-1)|x-y|}\right]  \nonumber \\
&&\bigotimes_{\alpha =2}^{N}\left| \Phi (0)\right\rangle _{\alpha }\int {\it %
D}\left[ \tilde{\varphi}\right] {\it D}\left[ \tilde{\varphi}^{\prime }%
\right] {\it D}\left[ \varphi _{1}\right] {\it D}\left[ \varphi _{1}^{\prime
}\right] \exp \frac{ic^{2}}{2\hslash }\int dtdx\left[ -\varphi _{1}\nabla
^{2}\varphi _{1}+\varphi _{1}^{\prime }\nabla ^{2}\varphi _{1}^{\prime }%
\right]  \nonumber \\
&&{\it T}\exp \left[ -i\frac{2mc}{\hslash }\sqrt{\frac{\pi G}{N-1}}\int
dtdx\left( \left[ \tilde{\varphi}(x)+\varphi _{1}(x)\right] \psi
_{1}^{\dagger }(x)\psi _{1}(x)-\tilde{\varphi}(x)g(x)\right) \right] \left|
\Phi (0)\right\rangle _{1}  \nonumber \\
&&_{1}\left\langle \Phi (0)\right| {\it T}^{-1}\exp \left[ i\frac{2mc}{%
\hslash }\sqrt{\frac{\pi G}{N-1}}\int dtdx\left( \left[ \tilde{\varphi}%
^{\prime }(x)+\varphi _{1}^{\prime }(x)\right] \psi _{1}^{\dagger }(x)\psi
_{1}(x)+\tilde{\varphi}^{\prime }(x)g^{\prime }(x)\right) \right]
\end{eqnarray}

In the equation above, the c-number factor inside the functional integration
on $g$, $g^{\prime }$ for large $N$ is the transition amplitude between the
evolved metastates respectively in the presence of external classical mass
densities proportional to $g$ and $g^{\prime }$. Now consider that, inside
the metastate space $S$ and for large $N$, the terms containing $g$ and $%
g^{\prime }$ in the double space integrals have vanishing mean square
deviations (as can be checked at any perturbative order), though their
commutators with the other terms are $O(N^{0})$, ($\tilde{\varphi},\tilde{%
\varphi}^{\prime }$ integrations imply $g,g^{\prime }\sim \psi ^{\dagger
}\psi $). The latter property forces us, when defining the expression
explicitly as the limit as $dt\rightarrow 0$ of time ordered products of
time evolution operators during time intervals $dt$, to keep the factors
depending on $g$ and $g^{\prime }$ in the right place, while the former one
allows for the replacement of the exponential of these terms with the
exponential of their average in the metastate at that time instant. As a
result, if we remember that $\left\langle \psi _{\alpha }^{\dagger }(x)\psi
_{\alpha }(x)\right\rangle =\left\langle \psi _{1}^{\dagger }(x)\psi
_{1}(x)\right\rangle $, we have 
\[
M(t)\mathrel{\mathop{\sim }\limits_{N\rightarrow \infty }} 
\]
\begin{eqnarray}
&&\int {\it D}\left[ \tilde{\varphi}\right] {\it D}\left[ \tilde{\varphi}%
^{\prime }\right] {\it D}\left[ g\right] {\it D}\left[ g^{\prime }\right]
\int {\it D}\left[ \varphi _{1}\right] {\it D}\left[ \varphi _{1}^{\prime }%
\right] \exp \frac{ic^{2}}{2\hslash }\int dtdx\left[ -\varphi _{1}\nabla
^{2}\varphi _{1}+\varphi _{1}^{\prime }\nabla ^{2}\varphi _{1}^{\prime }%
\right]  \nonumber \\
&&{\it T}\exp \frac{-i2mc}{\hslash }\sqrt{\frac{\pi G}{N}}\int dtdx 
\nonumber \\
&&\left( \left[ \tilde{\varphi}(x)+\varphi _{1}(x)\right] \psi _{1}^{\dagger
}(x)\psi _{1}(x)-g(x)\left[ \tilde{\varphi}(x)+\frac{m}{2c}\sqrt{\frac{GN}{%
\pi }}\int dy\frac{\left\langle \psi _{1}^{\dagger }(y)\psi
_{1}(y)\right\rangle }{|x-y|}\right] \right) \left| \Phi (0)\right\rangle
_{1}  \nonumber \\
&&_{1}\left\langle \Phi (0)\right| {\it T}^{-1}\exp i\frac{2mc}{\hslash }%
\sqrt{\frac{\pi G}{N}}\int dtdx  \nonumber \\
&&\left( \left[ \tilde{\varphi}^{\prime }(x)+\varphi _{1}^{\prime }(x)\right]
\psi _{1}^{\dagger }(x)\psi _{1}(x)-g^{\prime }(x)\left[ \tilde{\varphi}%
^{\prime }(x)+\frac{m}{2c}\sqrt{\frac{GN}{\pi }}\int dy\frac{\left\langle
\psi _{1}^{\dagger }(y)\psi _{1}(y)\right\rangle }{|x-y|}\right] \right) ,
\end{eqnarray}
where $\left\langle \psi _{1}^{\dagger }(y)\psi _{1}(y)\right\rangle \equiv
\left\langle \tilde{\Phi}(t)\right| \psi _{1}^{\dagger }(y,t)\psi
_{1}(y,t)\left| \tilde{\Phi}(t)\right\rangle $. By integrating over $%
g,g^{\prime },\tilde{\varphi},\tilde{\varphi}^{\prime }$, we get 
\[
M(t)\mathrel{\mathop{\sim }\limits_{N\rightarrow \infty }} 
\]
\begin{eqnarray*}
&&\int {\it D}\left[ \varphi _{1}\right] {\it D}\left[ \varphi _{1}^{\prime }%
\right] \exp \frac{ic^{2}}{2\hslash }\int dtdx\left[ -\varphi _{1}\nabla
^{2}\varphi _{1}+\varphi _{1}^{\prime }\nabla ^{2}\varphi _{1}^{\prime }%
\right] \\
&&{\it T}\exp \left[ \frac{iG}{\hslash }m^{2}\int dt\int dxdy\frac{\psi
^{\dagger }(x)\psi (x)\left\langle \psi ^{\dagger }(y)\psi (y)\right\rangle 
}{|x-y|}-\frac{i2mc}{\hslash }\sqrt{\frac{\pi G}{N}}\int dtdx\varphi
_{1}(x)\psi ^{\dagger }(x)\psi (x)\right] \left| \Phi (0)\right\rangle \\
&&\left\langle \Phi (0)\right| {\it T}^{-1}\exp \left[ \frac{-iG}{\hslash }%
m^{2}\int dt\int dxdy\frac{\psi ^{\dagger }(x)\psi (x)\left\langle \psi
^{\dagger }(y)\psi (y)\right\rangle }{|x-y|}+\frac{i2mc}{\hslash }\sqrt{%
\frac{\pi G}{N}}\int dtdx\varphi _{1}^{\prime }(x)\psi ^{\dagger }(x)\psi (x)%
\right]
\end{eqnarray*}
omitting the by now irrelevant index in $\psi _{1}$, and finally, after
integrating out $\varphi _{1},\varphi _{1}^{\prime }$, 
\[
\lim_{N\rightarrow \infty }M(t)\equiv \left| \Phi (t)\right\rangle
\left\langle \Phi (t)\right| = 
\]
\begin{eqnarray}
&&{\it T}\exp \left[ \frac{iG}{\hslash }m^{2}\sum_{\alpha <\beta }\int
dt\int dxdy\frac{\psi ^{\dagger }(x,t)\psi (x,t)\left\langle \Phi (t)\right|
\psi ^{\dagger }(y,t)\psi (y,t)\left| \Phi (t)\right\rangle }{|x-y|}\right]
\left| \Phi (0)\right\rangle  \nonumber \\
&&\left\langle \Phi (0)\right| {\it T}^{-1}\exp \left[ \frac{-iG}{\hslash }%
m^{2}\sum_{\alpha <\beta }\int dt\int dxdy\frac{\psi ^{\dagger }(x,t)\psi
(x,t)\left\langle \Phi (t)\right| \psi ^{\dagger }(y,t)\psi (y,t)\left| \Phi
(t)\right\rangle }{|x-y|}\right] ,  \label{SN}
\end{eqnarray}
where the normalization is automatically correct, as the resulting dynamics,
though nonlinear, is unitary. It should be remarked that, to make the
derivation more rigorous, the Newton potential should be replaced with \ a
regularized potential like $1/(|x-y|+\lambda )$ with $\lambda >0$, and the
Laplacian with the corresponding inverse. One should first take the limit $%
N\rightarrow \infty $, then take $dt\rightarrow 0$ and remove the
regularization, $\lambda \rightarrow 0$.

Eq. (\ref{SN}) coincides with the time evolution of the SN model in the
interaction representation, which is then the $N\rightarrow \infty $ limit
of our model. Just as in condensed matter physics, this limit suppresses
fluctuations (quantum fluctuations here) and preserves mean field features
only. In the present case the $N\rightarrow \infty $ limit keeps the
presence of localized states, but wipes out the non unitary evolution and
then the ability of generating decoherence.

In conclusion the present construction is the first derivation of the SN
model from a well defined quantum model producing both classical
gravitational interactions, localization and decoherence. In fact the model
was usually presented, up to now, as some sort of mean field approximation
of a not yet well specified theory incorporating self-interactions.

Acknowledgments - I would like to thank my colleague Mario Salerno for his
warm encouragement and stimulating discussions. Financial support from
M.U.R.S.T., Italy and I.N.F.M., Salerno is acknowledged

\end{document}